\documentclass{article}

 \usepackage[dblblindworkshop,final]{neurips_2025}

\workshoptitle{Frontiers in Probabilistic Inference: Sampling Meets Learning}

\usepackage[utf8]{inputenc} 
\usepackage[T1]{fontenc}    
\usepackage{hyperref}       
\usepackage{url}            
\usepackage{booktabs}       
\usepackage{amsfonts}       
\usepackage{nicefrac}       
\usepackage{microtype}      
\usepackage{xcolor}         
\usepackage{graphicx}  
\usepackage{algorithm} 
\usepackage{algpseudocode} 
\usepackage{float}
\newfloat{algorithm}{t}{lop} 
\usepackage{nicefrac}

\setcitestyle{round}

\title{Ising Machines for Model Predictive Path Integral-Based Optimal Control}

\author{%
  Lorin Werthen-Brabants \\
  imec - Ghent University\\
  9052 Ghent, Belgium \\
  \texttt{lorin.werthenbrabants@ugent.be} \\
  \And
  Pieter Simoens \\
  imec - Ghent University \\
  9052 Ghent, Belgium \\
  \texttt{pieter.simoens@ugent.be} \\
}

\usepackage{amsmath}

\begin{document}

\maketitle

\begin{abstract}
    We present a sampling-based Model Predictive Control (MPC) method that implements Model Predictive Path Integral (MPPI) as an \emph{Ising machine}, suitable for novel forms of probabilistic computing.
    By expressing the control problem as a Quadratic Unconstrained Binary Optimization (QUBO) problem, we map MPC onto an energy landscape suitable for Gibbs sampling from an Ising model. This formulation enables efficient exploration of (near-)optimal control trajectories. We demonstrate that the approach achieves accurate trajectory tracking compared to a reference MPPI implementation, highlighting the potential of Ising-based MPPI for real-time control in robotics and autonomous systems.
\end{abstract}

\section{Introduction}

Model Predictive Control (MPC) \citep{Camacho2007} is a widely used optimal control strategy, used to predict the next optimal action to be taken in robotics and automated driving, in which the optimal control input sequence $\{u_n\}$ is computed by solving a finite-horizon optimization problem at each time step. The optimization relies on a model of the system dynamics, $x_{n+1} = f(x_n,u_n)$, to predict future states and determine an optimal trajectory. Although MPC often yields good performance and robustness, the need to repeatedly solve an optimization problem in real time can make it computationally demanding, particularly for high-dimensional systems or those operating under strict energy or latency constraints.

Model Predictive Path Integral (MPPI) control \citep{williams2016aggressive} offers a probabilistic perspective on MPC by recasting the finite-horizon optimization as a stochastic sampling problem, rather than a direct deterministic optimizer, as is done when making use of Quadratic Programming (QP) optimizers such as SNOPT \citep{gill2005snopt} or Clarabel \citep{goulart2024clarabel}. Instead of solving a deterministic optimization problem directly, MPPI samples control trajectories from a distribution --- typically Gaussian --- and weights them according to the exponential of their negative cost, analogous to a Boltzmann distribution in statistical mechanics.

Recent advances in alternative computing paradigms provide new opportunities to accelerate computation. In particular, probabilistic bits (p-bits) \citep{camsari2017stochastic} have emerged as a promising technology for probabilistic computing. These p-bit based computers, or p-computers, are able to sample states from Ising machines and a Boltzmann distribution faster and more energy efficient than their CPU counterparts. FPGA simulations have been shown to provide approximately a 10$\times$ improvement in sampling throughput compared to highly optimized TPUs and GPUs \citep{chowdhury2023full}. Unlike deterministic bits, p-bits stochastically fluctuate between binary states, allowing them to efficiently explore energy landscapes described by Boltzmann distributions. This makes p-bits well suited for implementing MPPI in binary or finite action spaces, where control selection can be naturally mapped to energy-based models such as the Ising model.


Prior work has mapped MPC with finite inputs directly to QUBO and solved it via annealing-style optimizers \citep{inoue2020model}. Our contribution differs by recasting \textit{sampling-based} MPPI on a Boltzmann energy landscape and use Gibbs sampling to realize MPPI updates in binary spaces (Ising-MPPI), natively targeting probabilistic hardware (p-bits) \citep{si2024energy,kim2024fully}, rather than annealing based approaches such as with quantum annealing \citep{morita2008mathematical,inoue2020model}.

\section{Ising-based Model Predictive Path Integral Control}
Model Predictive Path Integral (MPPI) control \citep{williams2016aggressive} can be understood as sampling trajectories according to a Boltzmann distribution over their costs. In the standard \emph{continuous} setting, control sequences are perturbed by Gaussian noise vector $\mathbf{\epsilon} \sim \mathcal{N}(\mu,\sigma)$, and each noisy trajectory~$\mathbf{u} + \mathbf{\epsilon}$ is assigned a weight, where $J(\mathbf{u})$ is the trajectory cost, $w(\mathbf{\epsilon}) = \nicefrac{\exp\!\left(-\tfrac{1}{\lambda} J(\mathbf{u} + \mathbf{\epsilon})\right)}{Z}$, where $Z$ is the sum of all sampled $w(\epsilon)$, so that low-cost perturbations are exponentially more likely. This corresponds to sampling from a Boltzmann distribution, where the trajectory cost $J(\mathbf{u})$ plays the role of energy and the temperature parameter $\lambda$ regulates exploration.

However, an Ising machine does not work with continuous-valued inputs, but rather spin valued ($-1/\!+\!1$) or binary ($0/1$) inputs. In the binary setting the same principle applies as in the continuous one, but the action sequence is represented as a binary vector $\mathbf{a} \in \{0,1\}^d$, transformed to continuous quantities by calculating $E\mathbf{a}$, where $E$ is a binary expansion matrix \citep{inoue2020model} as described in Equation~\mbox{\ref{eqn:binary expansion}} . The Boltzmann distribution is then defined by a quadratic energy function, with $d=N\!Lm$, $N$ being the horizon length of the control problem, $L$ the number of bits for expansion matrix $E$ and $m$ the number of control inputs:

\begin{equation}\label{eqn:boltzmann}
p(\mathbf{a}) \propto \exp\!\left(-\tfrac{1}{\lambda} H(\mathbf{a})\right),  \quad \text{where } H(\mathbf{a}) = \mathbf{a}^\top \mathbf{J} \mathbf{a} + \mathbf{h}^\top \mathbf{a},\quad \mathbf{a} \in \{0,1\}^d.
\end{equation}

For the proposed Ising-MPPI technique, the construction of $\mathbf{J}$ and $\mathbf{h}$ is achieved by transforming a linear (or linearized) MPC problem into a QUBO formulation, as outlined in Appendix \ref{appendix:mpc}, specifically Equations \ref{eqn:interaction matrix} and \ref{eqn:bias}.

Instead of Gaussian perturbations, The proposed Ising-MPPI makes use of Gibbs sampling \citep{koller2009probabilistic,chowdhury2023full} to explore the binary action space, producing samples~$\{\mathbf{a}^{(s)}\}$ distributed according to the distribution in Equation \ref{eqn:boltzmann}. In probabilistic computers, this process is implemented efficiently and asynchronously \citep{chowdhury2023full}. For a binary vector~$\mathbf{a} \in \{0,1\}^d$, the conditional probability for bit~$i$ is computed from the local field~$p(a_i=1|a_{\setminus i}) = \sigma(-\nicefrac{1}{\lambda} ( h_i + 2\sum_{j \neq i} J_{ij} a_j ))$ with $\sigma$ the logistic function $\sigma(z_i) = \nicefrac{1}{1 + e^{-z_i}}$. For Ising-MPPI the empirical mean of these samples is computed over $S$ sweeps, and the final action is chosen by rounding each component $\hat{\mathbf{a}} = \mathbf{1}\!(\nicefrac{1}{S} \sum_{s=1}^S \mathbf{a}^{(s)} > 0.5)$, which can then be transformed to the actual action sequence $\hat{\mathbf{u}} = \mathbf{E}\hat{\mathbf{a}}$. This is done to anticipate hardware implementation where a long time constant $RC$ circuit could calculate the average voltage \citep{camsari2019p}, and could then be rounded to the nearest binary value. The entire procedure is outlined in Algorithm \ref{alg:imppi}.

At each control step --- MPPI outer iteration --- we re-linearize around the current state $x_t$ and the latest nominal control $\mathbf{\bar{u}}$, then rebuild $\mathbf{A}, \mathbf{B}, \mathbf{c}$ and $(\mathbf{J},\mathbf{h})$ accordingly. Just as MPPI converges to the optimal control by sampling and reweighting noisy perturbations according to a Boltzmann distribution over costs, the binary formulation converges to the optimal discrete control by Gibbs sampling from the corresponding Boltzmann distribution and rounding. 

\begin{algorithm}
\caption{Ising-based Model Predictive Path Integral (Ising-MPPI)}
\begin{algorithmic}[1]
\State \textbf{Input:} Reference $\mathbf{x}^{\text{ref}}$, horizon $N$, temperature $\lambda$, Gibbs samples $S$, MPPI Iterations $M$
\State $\mathbf{\bar{u}} \leftarrow \mathbf{0}$ \Comment{Initialize nominal control sequence}
\For{$1$ to $M$}
    \State $\mathbf{J},\mathbf{h} \leftarrow$ Equations \ref{eqn:interaction matrix} and \ref{eqn:bias}, using $\mathbf{x}^{\text{ref}}$ and horizon $N$ \Comment{Construct $\mathbf{J}, \mathbf{h}$ matrices}
    \State  $\{\mathbf{a}^{(s)}\} \leftarrow \text{Gibbs}(\mathbf{J, h}; S, \lambda)$  \Comment{Generate $S$ binary samples via Gibbs sampling}
    \State $\hat{\mathbf{a}} \leftarrow \mathbf{1}\left(\nicefrac{1}{S}\sum_{s=1}^S \mathbf{a}^{(s)} > 0.5\right)$ \Comment{Round to nearest binary value}
    \State $\bar{\mathbf{u}} \leftarrow \bar{\mathbf{u}} + \mathbf{E} \hat{\mathbf{a}}$ \Comment{Map to continuous actions}
\EndFor
\State Apply first control input $\mathbf{\bar{u}}_0$
\end{algorithmic}
\label{alg:imppi}
\end{algorithm}

\section{Results}\label{sec:results}

The Ising-MPPI approach is tested on a non-linear kinematic bicycle model, described in Appendix~\ref{appendix:bicycle model}. Three different MPC setups are tested: 1) the \textbf{Ising-MPPI} setting, following the approach outlined in Algorithm \ref{alg:imppi}, 2) a \textbf{Non-Ising Linear MPPI} setting, which makes use of the same $\mathbf{J}$ and $\mathbf{h}$ as the Ising-MPPI (Equations \ref{eqn:interaction matrix} and \ref{eqn:bias}), but with the $\mathbf{E}$ terms removed from their formulation, as explained in Section \ref{sec:nonising_linear_mppi}. The MPPI error is defined by the energy function $H$ in Equations \ref{eqn:boltzmann} and \ref{eqn:energy}, but applied on the action vector $\mathbf{u}$ rather than the binary version $\mathbf{a}$. In this setup, we can assess the effect of linearization outside of the binary-valued Ising context. Lastly, 3) a \textbf{reference} implementation making use directly of the nonlinear bicycle model, as introduced by \citet{williams2016aggressive}. The default hyperparameters for the experiments are listed in Table \ref{tab:hyperparams}.

All experiments were conducted on a workstation running Ubuntu 22.04.3 LTS, equipped with dual Intel Xeon Silver 4210 CPUs (40 threads total, 2.20 GHz) and 256 GB of RAM. All computations were performed on the CPU without GPU acceleration.


\subsection{Overall Performance}

We test the mean error over many different trajectories generated according to Appendix \ref{appendix:random trajectories}, with $M=4$, $S=200$ for Ising-MPPI, $M=4$, $S=1000$ for Non-Ising Linear MPPI and $M=4$, $S=1000$ for the reference MPPI. These values were found empirically to work well. We generate 50 different trajectories and run the MPPI algorithms in the three different setups with 10 different sampling seeds for a total of 500 trials.

The resulting tracking errors are shown in Table \ref{tab:tracking_error}. As expected, the reference setup achieves the lowest error with very small variance, serving as an ideal baseline. Ising-MPPI obtains a lower mean error than Non-Ising Linear MPPI, but its variance is much larger, indicating less consistent performance. In contrast, Non-Ising Linear MPPI shows slightly higher average error but with reduced variability across trials.

\begin{table}
\centering
\caption{Tracking error for the different setups. The tracking error is calculated as the mean squared error (MSE) between the predicted trajectory and the control trajectory. The standard deviation is also reported.\label{tab:tracking_error}}
\begin{tabular}{@{}ll@{}}
\toprule
\textbf{Type} & \textbf{Tracking Error (MSE)} \\ \midrule
Ising-MPPI   & 0.0271 $\pm$ 0.1085 \\
Non-Ising Linear MPPI & 0.0383 $\pm$ 0.0656 \\
Reference   & 0.0015 $\pm$ 0.0019 \\ \bottomrule
\end{tabular}
\end{table}

\subsection{Impact of Number of MPPI Samples on Trajectory Quality}\label{sec:samples_vs_quality}

\begin{figure}
    \centering
    \includegraphics[width=1.0\linewidth]{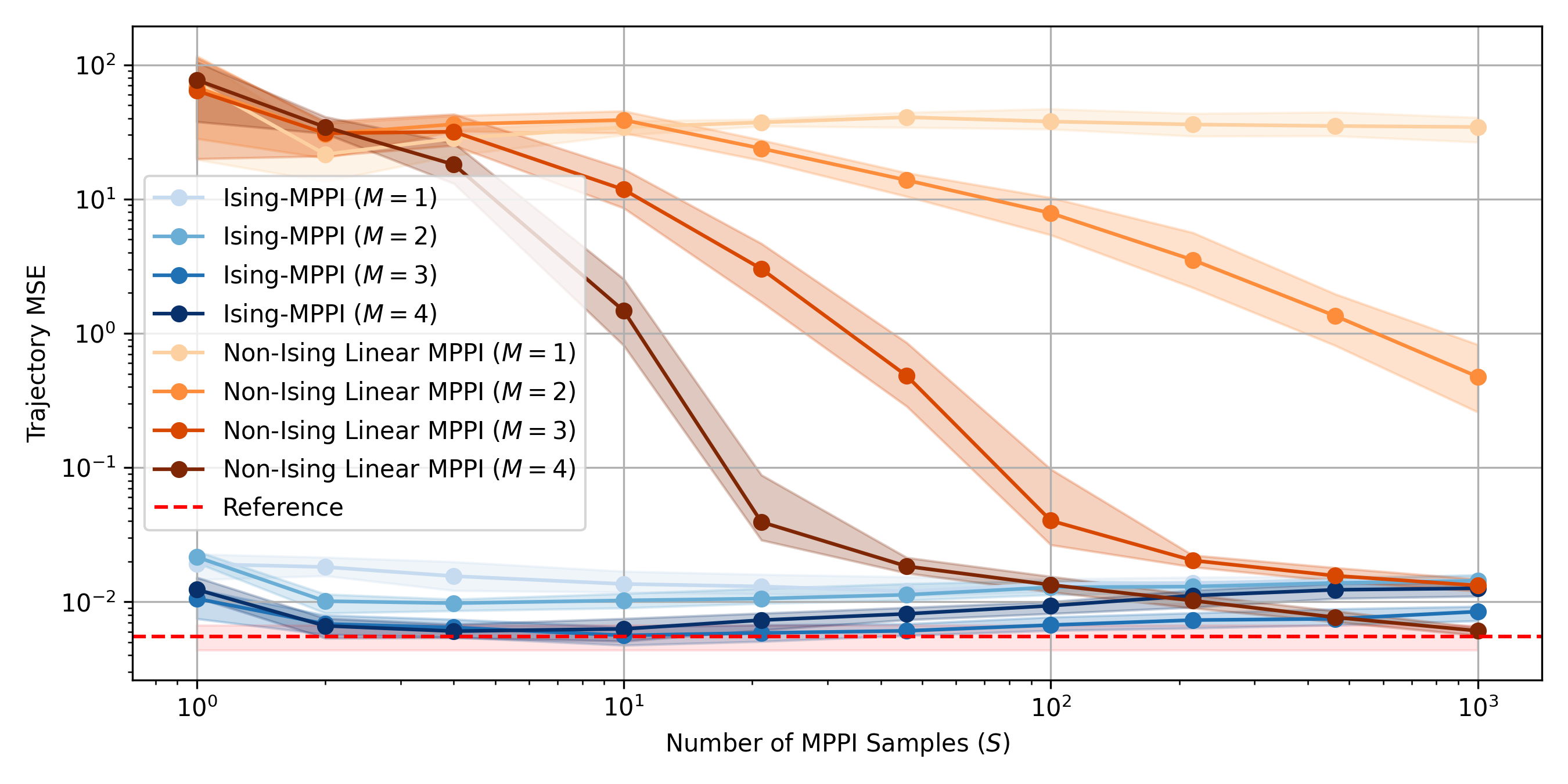}
    \caption{Comparison of Ising-MPPI vs Non-Ising Linear MPPI samples across different iteration limits. The $x$-axis represents the number of samples $S$ (log scale), the $y$-axis shows the MSE (log scale). Different shades of blue and orange denote a differing amount of iterations $M$ for the proposed finite-set Ising-MPPI and Non-Ising Linear MPPI respectively. The shaded areas represent the standard deviation. The red dashed line denotes the reference MPPI implementation.}
    \label{fig:continuous_vs_discrete}
\end{figure}

Figure~\ref{fig:continuous_vs_discrete} shows the effect of the number of MPPI samples on trajectory quality for a single, specific trajectory (Trajectory 1 from Figure \ref{fig:example_trajectories}), measured by the mean squared error (MSE). It can be seen that Ising-MPPI provides consistently lower MSE compared to the Non-Ising Linear MPPI formulation. The Non-Ising Linear MPPI converges at higher sample sizes ($S=10^3$) in a predictable way, with more iterations $M$ leading to lower errors with fewer samples per iteration $S$. It is important to highlight that the Non-Ising Linear MPPI cannot be solved on specialized Ising hardware and only serves to show the impact of linearization for MPPI. On the other hand the Ising-MPPI sampling approach converges to a low trajectory error faster. A surprising behaviour to be noted here is that the number of iterations for the discrete case does not simply reduce the error, but affects the error more unpredictably (e.g.\ $M=3$ having a lower MSE than $M=4$), unlike the Non-Ising Linear MPPI in which there is a clear inverse correlation between iterations $M$ and the tracking error. This is likely due to over-smoothing, occurring when averaging many binary control samples leads to values near $\nicefrac{1}{2}$, so rounding them flattens the control signal and reduces its responsiveness.

\section{Limitations}\label{sec:limitations}

First, the QUBO mapping relies on linearization around a nominal trajectory; in strongly nonlinear regimes or when the system departs significantly from the nominal path, the local linear approximation can become inaccurate, and iterative relinearization only partially mitigates this. Second, discretizing control inputs introduces quantization error: achieving very fine control resolution requires many binary variables, which increases the sampling and encoding cost. Third, our empirical validation here is limited to a kinematic bicycle model and CPU-based sampling: the hardware advantages we discuss need hardware-in-the-loop experiments with p-bit or FPGA Ising machines for full validation. Finally, the approach assumes reasonable model knowledge (dynamics and Jacobians); significant model error will degrade linearization quality and performance. 

\section{Conclusion}

In this work, we introduced a novel formulation of MPC with Ising Machine-based MPPI control. We show that MPC can be reinterpreted as sampling from a Boltzmann distribution over binary control sequences. This viewpoint bridges stochastic control and energy-based models, and allows utilization of specialized hardware for efficient trajectory sampling. The preliminary evaluation on a kinematic bicycle model shows similar results to the reference MPPI method, suggesting that binary sampling-based MPC can offer significant computational advantages when making use of p-computers.

From the previously discussed limitations in Section \ref{sec:limitations} we can derive several directions for future work. First, the current formulation assumes linearized dynamics around a nominal trajectory; extending the framework to handle more complex and less easily differentiable models and constraints would broaden its applicability, such as by encoding the model in a (deep) Boltzmann machine \citep{niazi2024training,salakhutdinov2009deep}. Second, hardware-in-the-loop experiments with FPGA- or actual p-bit-based Ising machines are necessary to validate the expected gains in sampling throughput and energy efficiency. Third, the discretization of control inputs introduces an approximation trade-off; exploring adaptive or multi-level quantization schemes could improve accuracy without sacrificing hardware compatibility.


\bibliography{library}

\clearpage

\appendix

\setcounter{equation}{0}
\renewcommand{\theequation}{A-\arabic{equation}}

\setcounter{table}{0}
\renewcommand{\thetable}{A-\arabic{table}}

\setcounter{figure}{0}
\renewcommand{\thefigure}{A-\arabic{figure}}

\section{MPC Formulation}\label{appendix:mpc}
\subsection{Linear MPC Formulation}

Consider a discrete-time system with linear dynamics \citep{Camacho2007}:

\begin{equation}
    x_{n+1} = A_n x_n + B_n u_n,
\end{equation}

where $x_n \in \mathbb{R}^s$ is the system state at time step $n$, $u_n \in \mathbb{R}^m$ is the control input, and $A_n$, $B_n$ are time-varying system matrices, with $s$ and $m$ the state and control dimensions. 

In the Model Predictive Control (MPC) framework, the objective is to compute a sequence of future control actions $\{u_n\}_{n=0}^{N-1}$ over a finite horizon $N$ that minimizes a cost function measuring the deviation of the system from a desired reference trajectory $\{x_n^{\text{ref}}\}$:

\begin{equation}
    \min_{\{u_n\}} \sum_{n=0}^{N-1} \mathcal{L}(x_n, x_n^{\text{ref}}) + \mathcal{R}(u_n),
\end{equation}

where $\mathcal{L}(x_n, x_n^{\text{ref}})$ quantifies state tracking error (e.g., quadratic error), and $\mathcal{R}(u_n)$ is a control regularization term to penalize excessive actuation. In a QUBO formulation, this can also be expressed as

\begin{equation}
J_n = x_n^\top Q x_n + u_n^\top R u_n,
\end{equation}

where $Q\in \mathbb{R}^s \succeq0$ and $R\in \mathbb{R}^m \succ 0$ are positive semidefinite and positive definite penalty matrices respectively.

At each time step, MPC solves this finite-horizon optimization using the current system state as the initial condition, applies the first control input $u_0$, and then repeats the process at the next time step. This receding-horizon strategy provides a balance between optimality and feedback robustness, enabling the controller to react dynamically to disturbances and model uncertainties.

\subsection{Finite set MPC as a QUBO formulation}

To reformulate Model Predictive Control (MPC) as a Quadratic Unconstrained Binary Optimization (QUBO) problem \citep{inoue2020model}, it is necessary to express the system dynamics over a finite prediction horizon in closed form. Consider a nonlinear system linearized around a sequence of nominal trajectories $(\mathbf{\bar{x}}, \mathbf{\bar{u}})$. For each time step $ n \in \{0, 1, \dots, N-1\} $, the system dynamics are approximated by:
\begin{equation}
x_{n+1} = (I + A_n\Delta t) x_n + B_n u_n \Delta t +(f(\bar{x}_n, \bar{u}_n) - A_n \bar{x}_n - B_n \bar{u}_n)\Delta t,
\end{equation}
where $ A_n $ and $ B_n $ are the Jacobians of the system dynamics with respect to the state and control inputs, respectively, evaluated at the nominal point $(\bar{x}_n, \bar{u}_n)$.

For sake of brevity, we will denote the time-discretized matrices $A_n\Delta t = A_n^\Delta$ and $B_n\Delta t = B_n^\Delta$. The block matrices $\mathbf{A}$ and $\mathbf{B}$, along with the constant vector $\mathbf{c}$, are constructed using the state transition matrices:
\begin{equation}
\Phi(i, j) =
\begin{cases}
\prod_{k=i}^{j-1} (I + A_k^\Delta), & \text{if } i < j, \\
I, & \text{if } i = j,
\end{cases}
\end{equation}
which describe the forward evolution of the linearized dynamics from time $i$ to $j$.

Using these definitions:
\begin{align}
    \mathbf{A} &= 
\begin{bmatrix}
\Phi(0,1) \\
\Phi(0,2) \\
\vdots \\
\Phi(0,N)
\end{bmatrix},\qquad
\mathbf{B} =
\begin{bmatrix}
0 & 0 & 0 & \cdots & 0 \\
\Phi(1,2) B_{1}^\Delta & 0 & 0 & \cdots & 0 \\
\Phi(1,3) B_{1}^\Delta & \Phi(2,3) B_{2}^\Delta & 0 & \cdots & 0 \\
\vdots & \vdots & \vdots & \ddots & \vdots \\
\Phi(1,N) B_{1}^\Delta & \Phi(2,N) B_{2}^\Delta & \Phi(3,N) B_{3}^\Delta & \cdots & 0
\end{bmatrix},\\
c_i &= \sum_{i=0}^{j-1} \Phi(i, j+1) \left[ f(\bar{x}_i, \bar{u}_i) - A_i \bar{x}_i - B_i \bar{u}_i \right]\Delta t.
\end{align}

The resulting block matrices $ \mathbf{A} $, $ \mathbf{B} $, and vector $ \mathbf{c} $ provide a compact, closed-form description of the system evolution over the prediction horizon:

\begin{equation}\label{eqn:closed form}
\begin{bmatrix}
x_1 \\
x_2 \\
\vdots \\
x_N
\end{bmatrix}
=
\underbrace{\mathbf{A} x_0}_{\text{initial state contribution}} +
\underbrace{\mathbf{B} \mathbf{u}}_{\text{control contribution}} +
\underbrace{\mathbf{c}}_{\text{nonlinear offset}},
\end{equation}
where, with $s$ the size of the state dimension, $m$ the size of the action dimension and $N$ the number of timesteps in the horizon:  $\mathbf{A} \in \mathbb{R}^{Ns \times s}$ captures the effect of the initial state, $\mathbf{B} \in \mathbb{R}^{Ns \times Nm}$ maps the stacked control inputs to the state trajectory, $\mathbf{c} \in \mathbb{R}^{Ns}$ includes constant terms arising from deviations from the linearization point.

\subsection{Binary Expansion Matrix}

To discretize continuous controls into binary form, we use an expansion
matrix $E \in \mathbb{R}^{m \times L}$, where~$m$ is the number of
control inputs and $L$ the number of binary variables per input. The
matrix is constructed with scaled powers of two:
\begin{align}\label{eqn:binary expansion}
E_{j,i} = K \frac{2^{i-1}}{2^{L-1}} \quad \text{for } j \in \{1,\dots,m\}, i \in \{1,\dots,L\},
\end{align}
with the sign of the most significant bit $E_{j,L} = -K$ to allow negative
values. Each row of $E$ is then repeated for all inputs to become a block matrix $\mathbf{E} \in \mathbb{R}^{Nm\times NL}$ where $N$ is the horizon length and $\mathbf{E}\mathbf{a} = \mathbf{u}, \mathbf{a} \in \{0,1\}^{NLm}, \mathbf{u} \in \mathbb{R}^{Nm}$. The hyperparameters used for experiments are listed in Table \ref{tab:hyperparams}.

\subsection{Interaction terms and bias}

Substituting the closed-form expression of the state trajectory into the cost function, and expressing the control input vector $\delta\mathbf{u} = E \mathbf{a}$, where $\mathbf{a}$ is a binary vector encoding the control actions, we obtain a quadratic expression of the following form, analogous to the approach by \citet{inoue2020model}:
\begin{equation}\label{eqn:energy}
H(a) = \mathbf{a}^\top \mathbf{J}\mathbf{a} + \mathbf{h}^\top \mathbf{a},
\end{equation}
where, with $x_0$ the current state and $\mathbf{\bar{u}}$ a nominal action vector
\begin{align}
\mathbf{J} &= \mathbf{E}^\top \left(\mathbf{B}^\top \mathbf{Q} \mathbf{B} + \mathbf{R}\right) \mathbf{E},\label{eqn:interaction matrix} \\
\mathbf{h}^\top &= 2 \left(\mathbf{A} x_0 + \mathbf{B}\bar{\mathbf{u}} + \mathbf{c} - \mathbf{x}^\text{ref}\right)^\top \mathbf{Q} \mathbf{B} \mathbf{E}.\label{eqn:bias}
\end{align}

Here, $\mathbf{A}$, $\mathbf{B}$, and $\mathbf{c}$ are derived from the closed-form representation discussed previously. $\mathbf{Q} \in \mathbb{R}^{Ns}$ and $\mathbf{R} \in \mathbb{R}^{Nm}$ are stacked penalty matrices $Q$ and $R$ respectively. $\mathbf{x}^\text{ref} \in \mathbb{R}^{Ns}$ is the reference trajectory that needs to be followed. The binary vector $\mathbf{a}$ parametrizes the control sequence via a linear mapping $ \mathbf{u} = \mathbf{E a} $, where $\mathbf{E}$ is the binary expansion block matrix. Note that in Equation \ref{eqn:bias} the effect of the nominal action sequence $\mathbf{B\bar{u}}$ is present. As a result, the optimal vectors $\delta\mathbf{u} = \mathbf{E a}$ are only deviations from the nominal action sequence $\mathbf{\bar{u}}$, in order to make it compatible with the MPPI formulation.

The matrix $\mathbf{J}$ defines the quadratic coefficients of the QUBO problem, while the vector $\mathbf{h}$ defines the linear coefficients. Together, these specify the optimization objective in standard QUBO form:
\begin{equation}
\min_{\mathbf{a} \in \{0,1\}^d} H(\mathbf{a})
\end{equation}
where $d$ is the total number of binary decision variables, in our case $d=NLm$.

\subsubsection{Symmetrization}

For a standard Ising formulation, $\mathbf{J}$ needs to be symmetric and have a zero-valued diagonal. In order to ensure this, we can apply the following method. Given a quadratic coefficient matrix $\mathbf{J}$ and linear vector $\mathbf{h}$, we apply the following preprocessing routine:
\begin{align}
\mathbf{J} &\leftarrow \tfrac{1}{2}(\mathbf{J} + \mathbf{J}^\top) &\text{(symmetrization)}, \\
\mathbf{h} &\leftarrow \mathbf{h} + \mathrm{diag}(\mathbf{J}) &\text{(absorb diagonal into biases)}, \\
\mathrm{diag}(\mathbf{J}) &\leftarrow 0 &\text{(remove self-interactions)}.
\end{align}

\subsection{Non-Ising Linear MPPI formulation}\label{sec:nonising_linear_mppi}

In order to analyze the impact of discretization and direct sampling of the Ising Machine, a \textbf{Non-Ising Linear MPPI} formulation is made, where the regular MPPI algorithm is applied with the Ising energy objective with the binary expansion block matrix $\mathbf{E}$ removed:

\begin{align}
\mathbf{J} &=  \mathbf{B}^\top \mathbf{Q} \mathbf{B} + \mathbf{R},\\
\mathbf{h}^\top &= 2 \left(\mathbf{A} x_0 + \mathbf{B}\bar{\mathbf{u}} + \mathbf{c} - \mathbf{x}^\text{ref}\right)^\top \mathbf{Q} \mathbf{B},
\end{align}

with Equation \ref{eqn:energy} ($H(a) = \mathbf{a}^\top \mathbf{J}\mathbf{a} + \mathbf{h}^\top \mathbf{a}$) used as the trajectory cost for MPPI.

\section{Random Trajectory Generation}\label{appendix:random trajectories}

To test the proposed binary MPPI controller, we require a variety of reference trajectories that reflect different driving conditions. For this purpose, we generate randomized spline-based trajectories using the procedure below.

The generator begins from a fixed start position at the origin, $(0,0)$, and iteratively samples a sequence of control points. At each step, a random heading angle and a random step distance (between 4 and 5 units) are drawn, and the new control point is placed relative to the previous one. The first heading angle is uniformly sampled in $[0, 2\pi)$, while subsequent headings are restricted to lie within a $\pm \pi/2$ cone around the previous segment direction. This ensures smooth trajectories without sharp reversals. The procedure is repeated until a total of eight control points are obtained.

\begin{figure}
    \centering
    \includegraphics[width=1.0\linewidth]{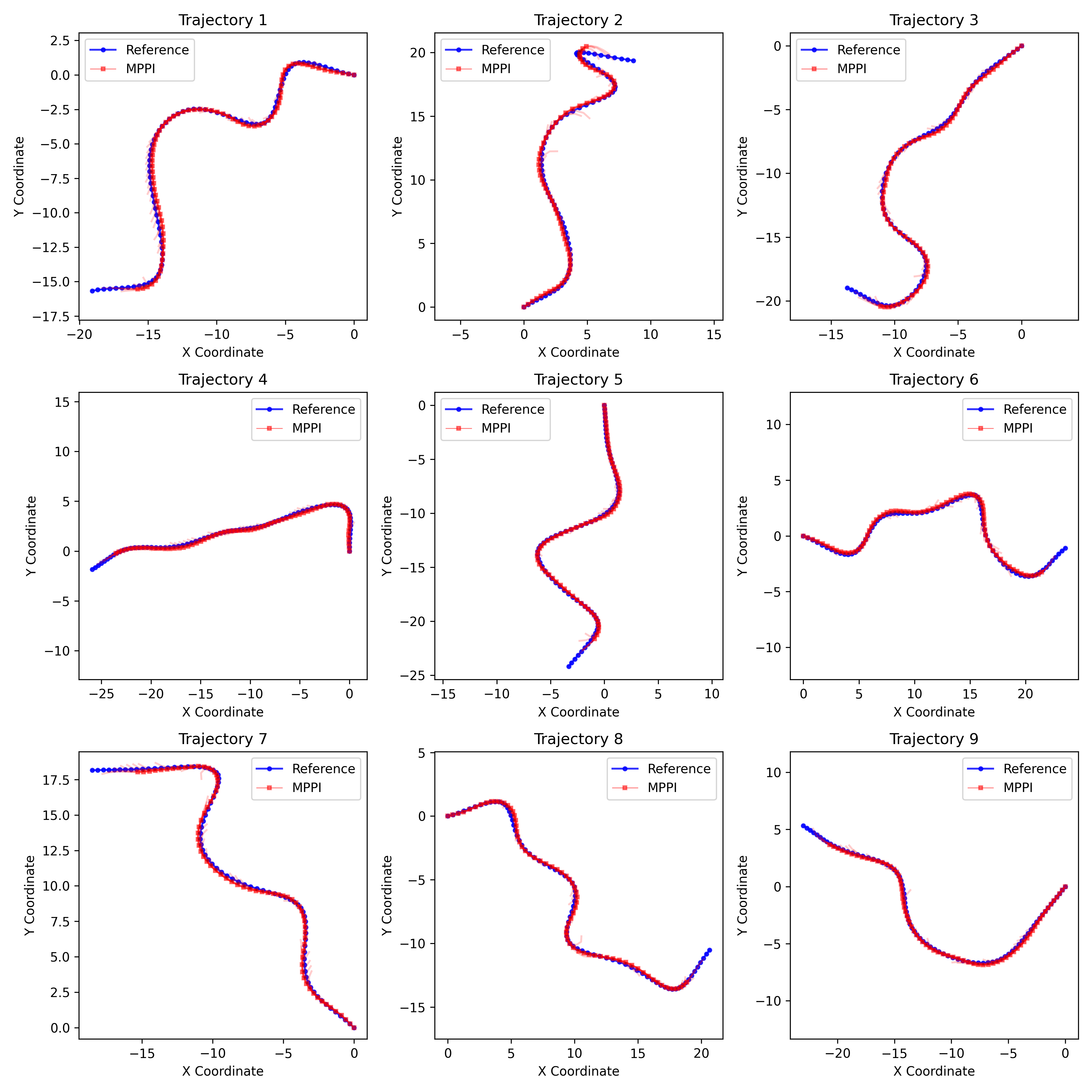}
    \caption{Examples of trajectories and their Ising-MPPI solutions. Note that the $N=8$ last points in the reference trajectory are not solved due to no additional reference points being available. Each trajectory starts at $(0, 0)$. The light red streaks visible are the predicted trajectories at each point.}
    \label{fig:example_trajectories}
\end{figure}

Examples of the proposed MPPI technique applied to different, randomly generated trajectories can be seen in Figure \ref{fig:example_trajectories}.

\section{Bicycle Model Benchmark Environment}\label{appendix:bicycle model}

To evaluate the proposed binary MPPI framework, we consider a simple kinematic bicycle model with steering dynamics. The system state and control inputs are defined as
\begin{equation}
x = [p_x,\, p_y,\, \theta,\, v,\, \delta]^\top, \quad
u = [a,\, \omega_\delta]^\top,
\end{equation}
where $(p_x, p_y)$ denote the vehicle position, $\theta$ is the heading angle, $v$ is the forward velocity, and $\delta$ is the steering angle. The control vector consists of the longitudinal acceleration $a$ and the steering rate $\omega_\delta$. The continuous-time dynamics \citep{de2005feedback} are given by 

\begin{align}
\begin{bmatrix}
\dot{p}_x \\
\dot{p}_y \\
\dot{\theta} \\
\dot{v} \\
\dot{\delta}
\end{bmatrix}
&=
\begin{bmatrix}
v \cos \theta \\
v \sin \theta \\
\frac{v}{L} \tan \delta \\
a \\
\omega_\delta
\end{bmatrix}.
\end{align}

For our tests, we set $L=1$. The system dynamics $f(x,u)$ are linearized around the current state $x$ and control input $u$. This is done by computing the Jacobian matrices of the dynamics with respect to the state and input:

\begin{align}
    \mathbf{f}_x = \frac{\partial f}{\partial x}, \qquad
    \mathbf{f}_u = \frac{\partial f}{\partial u},
\end{align}

where $\mathbf{f}_x$ represents the sensitivity of the system dynamics to the current state, and $\mathbf{f}_u$ represents the sensitivity to the control input.

Evaluating these Jacobians at a specific state $x$ and control $u$ gives the linearized system matrices:

\begin{align}
    A = \mathbf{f}_x \big|_{x,u}, \qquad
    B = \mathbf{f}_u \big|_{x,u}.
\end{align}

The matrices $A$ and $B$ provide a local linear approximation of the nonlinear dynamics around the chosen operating point, which is essential for control design and analysis techniques that rely on linear models.

\section{Hyperparameters}

\begin{table}[H]
\centering
\caption{Hyperparameters used in experiments.\label{tab:hyperparams}}
\begin{tabular}{@{}lll@{}}
\toprule
Category & Symbol / Name & Value \\ \midrule
\textbf{MPC/Dynamics} 
 & Horizon length & $N=8$ \\
 & Time step & $\Delta t = 0.1$ \\
 & Vehicle model & Kinematic bicycle \\
 \midrule
\textbf{Cost function} 
 & State cost $Q$ & $\mathrm{diag}([1000,1000,1,0,0])$ \\
 & Control cost $R$ & $\mathrm{diag}([1,1])$ \\
 \midrule
\textbf{Binary encoding} 
 & Bits per control input & $L=5$ \\
 & Magnitude of speed control & $K = 15$ \\
 & Magnitude of steering control & $K=2.2$ \\ \midrule
\textbf{Sampling (Gibbs / Ising)} 
 & Temperature & $\lambda = 0.1$ \\
 & Outer iterations & $M=4$ \\
 & Gibbs sweeps per sample & $S=1000$ \\
 \bottomrule
\end{tabular}
\end{table}

\end{document}